\documentclass[prl,aps,twocolumn,showpacs,superscriptaddress,amsfonts,amsmath,floatfix]{revtex4}

\usepackage{graphicx}
\usepackage{color}
\usepackage{import}
\usepackage{wrapfig}
\usepackage{mathtools}
\usepackage{bm}
\usepackage{amsmath}
\usepackage{amsfonts}
\usepackage{amssymb}

\begin{document}


\title{Dynamically Emerging Topological Phase Transitions \\ in Nonlinear Interacting Soliton Lattices}

\author{Domenico Bongiovanni}
\affiliation{The MOE Key Laboratory of Weak-Light Nonlinear Photonics, 
TEDA Applied Physics Institute and School of Physics, Nankai University, Tianjin 300457, China}
\affiliation{INRS-EMT, 1650 Blvd. Lionel-Boulet, Varennes, Quebec J3X 1S2, Canada}

\author{Dario Juki\'{c}}
\affiliation{Faculty of Civil Engineering, University of Zagreb, A. Ka\v{c}i\'{c}a  Mio\v{s}i\'{c}a 26, 10000 Zagreb, Croatia}

\author{Zhichan Hu}
\affiliation{The MOE Key Laboratory of Weak-Light Nonlinear Photonics, 
TEDA Applied Physics Institute and School of Physics, Nankai University, Tianjin 300457, China}

\author{Frane Luni\'{c}}
\affiliation{Department of Physics, Faculty of Science, University of Zagreb, Bijeni\v{c}ka c. 32, 10000 Zagreb, Croatia}

\author{Yi Hu}
\affiliation{The MOE Key Laboratory of Weak-Light Nonlinear Photonics, 
TEDA Applied Physics Institute and School of Physics, Nankai University, Tianjin 300457, China}

\author{Daohong Song}
\affiliation{The MOE Key Laboratory of Weak-Light Nonlinear Photonics, 
TEDA Applied Physics Institute and School of Physics, Nankai University, Tianjin 300457, China}

\author{Roberto Morandotti}
\affiliation{INRS-EMT, 1650 Blvd. Lionel-Boulet, Varennes, Quebec J3X 1S2, Canada}
\affiliation{Institute of Fundamental and Frontier Sciences, University of Electronic Science and Technology of China, Chengdu, Sichuan 610054, China}

\author{Zhigang Chen}
\email{zgchen@nankai.edu.cn }
\affiliation{The MOE Key Laboratory of Weak-Light Nonlinear Photonics, 
TEDA Applied Physics Institute and School of Physics, Nankai University, Tianjin 300457, China}
\affiliation{Department of Physics and Astronomy, San Francisco State University, San Francisco, California 94132, USA}

\author{Hrvoje Buljan}
\email{hbuljan@phy.hr}
\affiliation{The MOE Key Laboratory of Weak-Light Nonlinear Photonics, 
TEDA Applied Physics Institute and School of Physics, Nankai University, Tianjin 300457, China}
\affiliation{Department of Physics, Faculty of Science, University of Zagreb, Bijeni\v{c}ka c. 32, 10000 Zagreb, Croatia}

\date{\today}

\begin{abstract}
We demonstrate dynamical topological phase transitions in evolving Su-Schrieffer-Heeger (SSH) lattices made of 
interacting soliton arrays, which are {\it entirely} driven by nonlinearity and thereby exemplify emergent 
nonlinear topological phenomena. 
The phase transitions occur from topologically {\it trivial-to-nontrivial} phase in periodic succession with 
crossovers from topologically {\it nontrivial-to-trivial} regime. The signature of phase transition is 
gap-closing and re-opening point, where two extended states are pulled from the bands into the gap to become localized topological edge states. 
Crossovers occur via decoupling of the edge states from the bulk of the lattice.
\end{abstract}

\maketitle

Topological photonics offers a unique path for manufacturing photonic devices 
immune to scattering losses and disorder~\cite{Lu2014, Ozawa2019}. 
Since the pioneering theoretical predictions~\cite{Haldane2008} and 
experimental demonstrations~\cite{Wang2009} of topologically protected electromagnetic edge 
states, most studies have focused on linear topological photonic structures~\cite{Lu2014, Ozawa2019}. 
However, by combining topology with nonlinearity~\cite{Lumer2013, Ablowitz2014, Leykam2016, 
Solnyshkov2017, Dobrykh2018, Mukherjee2020, Xia2020, Bisianov2019, Hadad2016, Zhou2017, Maczewsky2020, Katan2016, 
Kruk2019, Wang2019,  Zhang2019, Malkova2009}, 
many opportunities for fundamental discoveries and new functionalities of the devices arise~\cite{Smirnova2020}; 
this is appealing also because nonlinearity inherently exists or is straightforwardly activated in most 
of the currently used linear topological photonic systems. 
The studies of nonlinear topological phenomena in photonics include, for example, nonlinear topological 
edge states and solitons~\cite{Lumer2013, Ablowitz2014, Leykam2016, Solnyshkov2017, Mukherjee2020, Zhang2019,
Dobrykh2018, Xia2020, Bisianov2019, Malkova2009}, 
topological phase transitions activated via nonlinearity~\cite{Hadad2016, Zhou2017, Maczewsky2020, Katan2016}, 
nonlinear frequency conversion~\cite{Kruk2019, Wang2019}, 
topological lasing~\cite{StJean2017, Bahari2017, Bandres2018, Harari2018, Zhao2018, Parto2018}, 
and nonlinear tuning of non-Hermitian topological states~\cite{XiaPT2020, Pernet2020}.

In a recent study, we have introduced the concepts of inherited and emergent nonlinear topological 
phenomena~\cite{Xia2020}. In this classification, inherited phenomena occur when nonlinearity is a 
small perturbation on an otherwise linear topological system. 
For example, in the SSH lattice~\cite{SSH}, nonlinearity can easily break the chiral symmetry and therefore the underlying topology; 
this enables coupling into an otherwise topologically protected edge state~\cite{Xia2020, Bisianov2019}.
However, many of the system properties, such as the structure of the nonlinear topological edge 
states and/or solitons~\cite{Lumer2013, Ablowitz2014, Leykam2016, Solnyshkov2017, Mukherjee2020, Zhang2019, Dobrykh2018, Xia2020, Bisianov2019}, are inherited from the corresponding linear system~\cite{Xia2020}.
In contradistinction, emergent nonlinear topological phenomena occur when the underlying linear system is not topological, 
but the nonlinearity induces nontrivial topology~\cite{Xia2020}. 
Nonlinearity induced topological phase transitions~\cite{Hadad2016, Zhou2017, Maczewsky2020, Katan2016}
are examples of emergent nonlinear topological phenomena. 
In a recent experiment utilizing a nonlinear waveguide lattice structure~\cite{Maczewsky2020}, 
such a transition was shown to happen when power exceeded a certain threshold value. 
Emergent nonlinear topological phenomena are intriguing but were scarcely explored in nonlinear topological 
photonics.

\begin{figure}
\includegraphics[width=0.45\textwidth]{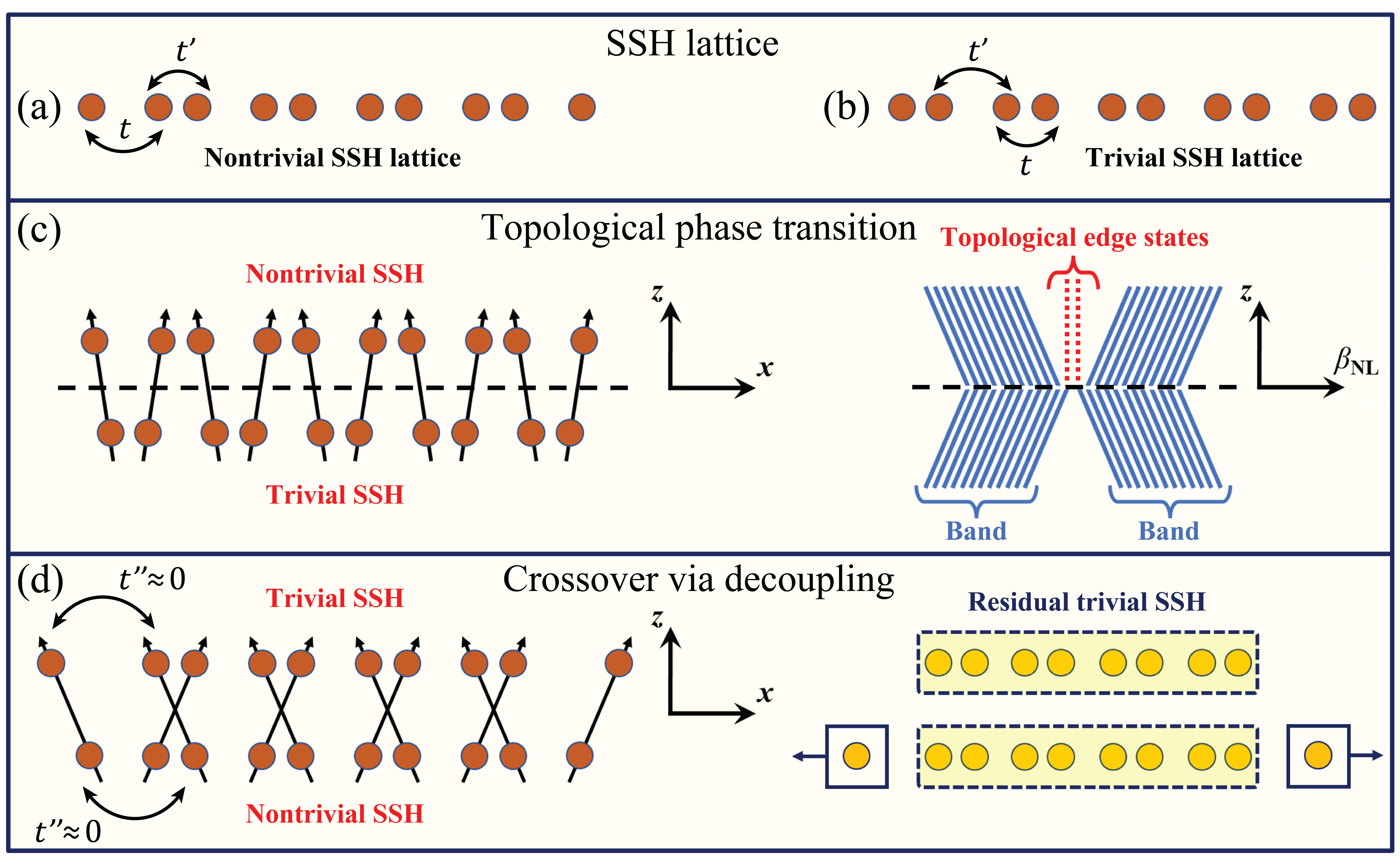}
\caption{Illustration of topological phase transition and crossover 
found in the evolving SSH soliton lattice. 
(a) The SSH lattice in the topologically nontrivial regime with $t<t'$, characterized by 
two localized topological edge states.
(b) The SSH lattice in the topologically trivial regime with $t>t'$. 
(c) Sketch of the topological phase transition from trivial-to-nontrivial 
phase in real space (left) and in the spectrum (right). 
At the phase transition, the gap closes, and two extended eigenmodes are pulled from the 
bands into the gap to become topological edge states. 
(d) Sketch of the crossover from the nontrivial-to-trivial phase via decoupling of the outermost lattice sites. 
The next-nearest-neighbor coupling is negligible in our SSH lattice, $t^{''}\approx 0$, 
which results in decoupling during evolution in our system (left). 
This is equivalent to pulling off the outermost SSH lattice sites to infinity,  
leaving the residual lattice in the trivial phase (right).
}
\label{fig:1}
\end{figure}

Here we report the dynamical topological phase transitions entirely driven by nonlinearity, which constitute an example of emergent nonlinear topological phenomena. These phase transitions occur in colliding soliton lattices and are enabled by elastic soliton collisions. In optics, spatial solitons are stable localized optical beams, which occur when diffraction is balanced by nonlinearity~\cite{Chen2012}. 
Here we create two one-dimensional (1D) soliton sublattices, and initially kick them in opposite directions. As the sublattices evolve and collide, their superposition forms a paradigmatic model of topological physics: 
the SSH lattice~\cite{SSH}, which can be in the topologically nontrivial [Fig.~\ref{fig:1}(a)] as well as trivial [Fig.~\ref{fig:1}(b)] 
phase featuring the so-called Zak phase~\cite{Zak1989}. 
We find two kinds of interesting phenomena, which periodically occur in succession: 
(i) a dynamical topological phase transition from topologically trivial-to-nontrivial phase, 
characterized by a gap closing and re-opening at a single point, where two extended states 
are pulled from the bands into the gap to become localized topological edge states [see Fig.~\ref{fig:1}(c)];
(ii)  a crossover from the topologically nontrivial-to-trivial regime, which occurs via decoupling 
of the edge states from the bulk of the lattice [see Fig.~\ref{fig:1}(d)].

We emphasize up front  that there is a distinction between our system, and those from 
Refs.~\cite{Hadad2016, Zhou2017, Maczewsky2020}, which all exhibit nonlinearity-induced topological phase transitions. 
In the theoretical models of Refs.~\cite{Hadad2016, Zhou2017}, the photonic lattices are fixed in $x$-space. 
In Ref.~\cite{Maczewsky2020} they are fixed in the $x$-$z$-space (i.e., ''spacetime''); the 
power of an external excitation can change the coupling via nonlinearity to induce a phase transition.  
In our system, the whole lattice autonomously nonlinearly evolves in the $x$-$z$-space, 
resulting in different topological phases along $z$ (i.e., ''time'').
The surprising connection between interacting soliton lattices and nontrivial topology is revealed by 
the phase transitions and crossovers accompanied by the ''birth'' and ''death'' of topological edge states. 
This reminds of the surprising connection between topology and quasicrystals, also revealed by the 
phase transitions~\cite{Kraus2012}.

We first outline a few basic facts about the SSH lattice. 
It is a 1D topological system, which exists due to the underlying chiral symmetry~\cite{SSH, Ozawa2019}. 
In its topologically nontrivial phase, the intercell coupling $t'$ is stronger than the intracell coupling $t$ ($t<t'$) 
[see Fig.~\ref{fig:1}(a)]. 
The nontrivial SSH lattice has two topological edge modes with propagation 
constants residing in the band gap and a characteristic phase structure~\cite{SSH, MalkovaOL2009}.
In the trivial phase $t>t'$ [see Fig. \ref{fig:1}(b)], there are two bands separated by a gap, and all 
eigenmodes are extended.  
This model has been implemented in versatile systems, including photonics and 
nanophotonics~\cite{MalkovaOL2009, Keil2013, Xiao2014, Blanco-Redondo2016, Weimann2017}, 
plasmonics~\cite{Poddubny2014, Bleckmann2017}, as well as 
quantum optics~\cite{Kitagawa2012, Cardano2017, Blanco-Redondo2018, Bello2014}. 
Some of the aforementioned nonlinear topological phenomena have been studied also in 
the nonlinear SSH model~\cite{Malkova2009, Xia2020, Kruk2019, Hadad2016, 
Solnyshkov2017, Zhou2017, Dobrykh2018, Bisianov2019, Parto2018, Zhao2018}.

We  consider the propagation of a linearly-polarized optical beam in a nonlinear medium, 
which is described by a nonlinear Schr{\"o}dinger equation (NLSE),
\begin{equation}
i \frac{\partial \psi}{\partial z}+\frac{1}{2k} \frac{\partial^2 \psi}{\partial x^2}+\gamma |\psi|^2 \psi(x,z)=0,
\label{NLSE}
\end{equation}
where $\psi(x,z)$ refers to the electric field envelope, $\gamma$ defines the strength of the nonlinearity 
(we assume a Kerr-type nonlinearity), and $k$ is the wavenumber in the medium. 
The NLSE possesses a family of soliton solutions, with the hyperbolic-secant soliton being the most 
representative~\cite{Agrawal-book}:
\begin{align}
\psi_S(x , z; \kappa, \theta)  = 
\sqrt{I_0}\,\mbox{sech} \left( \frac{x}{x_0} - \frac{\kappa z}{k x_0^2} \right)  \nonumber \\
\times \mbox{exp}\left[ i \left(\frac{\kappa}{x_0}x + \frac{1-\kappa^2}{2 k x_0^2} z+\theta
\right) \right].
\label{sechsol}
\end{align}
Here, $x_0$ is a scaling factor, $\kappa/x_0$ is the initial momentum, $I_0$ defines the peak intensity, and 
$\theta$ is an arbitrary phase. 
The stationary propagation is achieved when diffraction (quantified by the diffraction length $k x_0^2$) is 
balanced by the nonlinearity (quantified by the nonlinear length $1/\gamma I_0$), that is, when 
$\gamma I_0=(k x_0^2)^{-1}$. 

Systems described by Eq.~\eqref{NLSE} are usually unrelated to nontrivial topology.  
Such scenario emerges from initial condition(s) given by 
\begin{eqnarray}
\psi(x,0) & = & \sum_{j=-M}^M
\psi_S(x-T-jd, 0; -\kappa, 0) \nonumber \\
& + & \sum_{j=-M}^M
\psi_S(x+T-jd , 0; \kappa, \theta),
\end{eqnarray}
where the first sum relates to sublattice B, and the second to sublattice A. 
The parameter $d$ defines the size of the unit cell, and 
$T$ is the initial offset between the two sublattices. 
The next-nearest-neighbor (NNN) tunneling in our SSH lattice is negligible, $t^{''}\approx 0$.  
Due to the presence of nonlinearity, soliton interaction results in a dynamically 
evolving optically-induced lattice. To study its properties, 
we study the eigenvalues $\beta_{NL,n}(z)$ and the eigenmodes $\phi_{NL,n}(x,z)$ of the 
(nonlinearly) optically induced lattice potential $V(x,z)=-\gamma |\psi(x,z)|^2$, defined by 
$H\phi_{NL,n}=\beta_{NL,n}\phi_{NL,n}$; here $H=-(2k)^{-1}\partial_{xx}+V$. 
An equivalent approach for evolving nonlinear topological lattices was adopted in Ref. \cite{Xia2020}.

\begin{figure}
\includegraphics[width=0.50\textwidth]{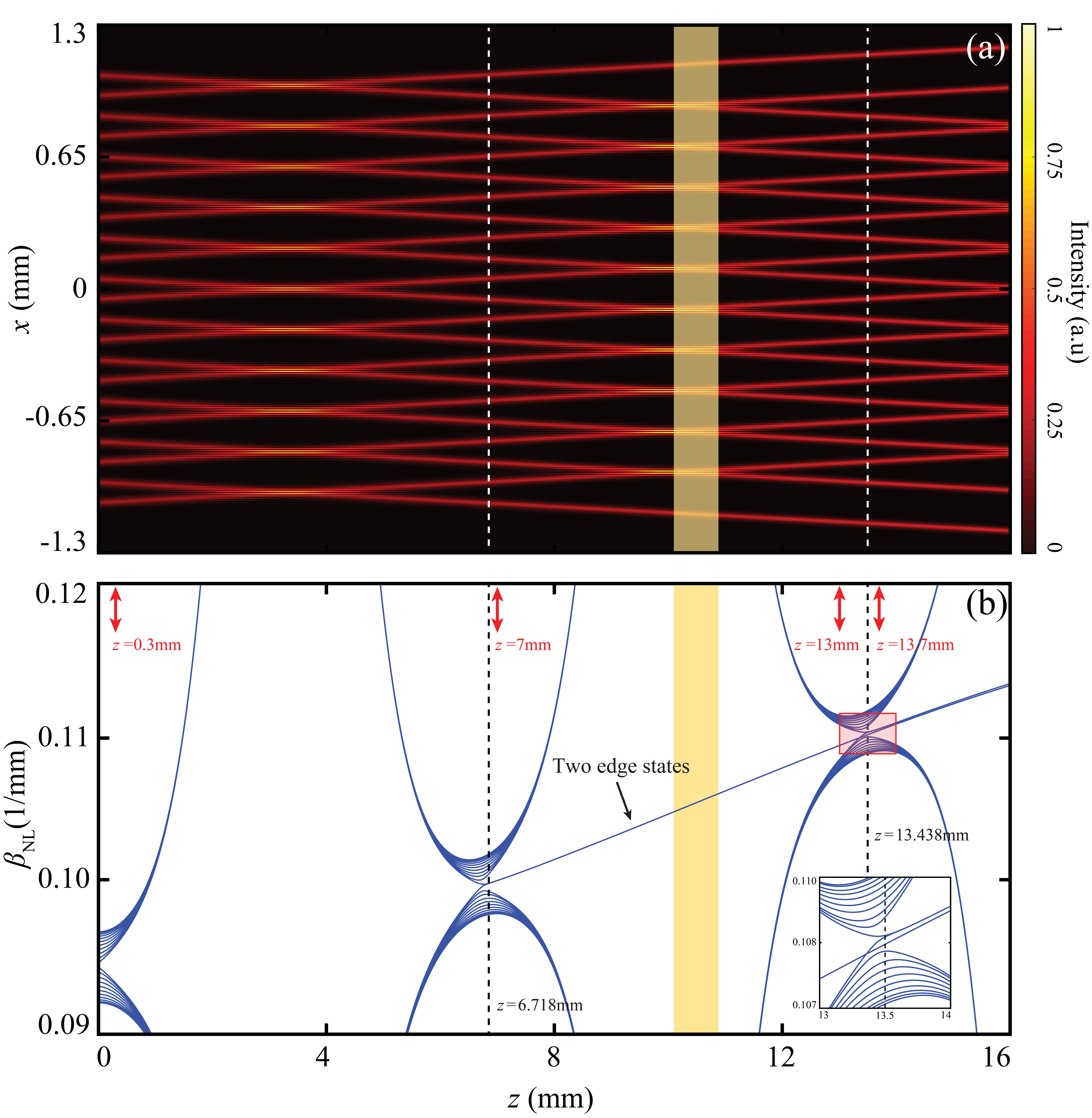}
\caption{
Intensity (a) and spectrum (b) of the SSH soliton lattice evolving with propagation distance $z$. 
Locations where topological phase transitions happen are indicated with vertical dashed lines, 
while crossover region is highlighted by the yellow stripe. Topological phase transitions occur 
at the gap-closing points, after which two extended eigenmodes are pulled from the bands into 
the gap and become topological edge states (i.e., the phase transition here is from trivial-to-nontrivial phase). 
Between these closing points, there is a crossover from nontrivial-to-trivial phase via 
decoupling of the edge states from the bulk of the lattice, which can be understood by comparing 
(a) with Fig.~\ref{fig:1}(d). At the second transition stage, two new edge states emerge, as seen clearly 
in the zoom-in inset in (b), while the old ones turn into decoupled walk-off solitons. 
Parameters: $M=5$, $T=d/4=50$~$\mu$m, $\theta=\pi$, $x_0 = 18.0~\mu$m, $\kappa = 5$, $k=1.71\times 10^{7}$~m$^{-1}$, 
and $\gamma I_0=(k x_0^2)^{-1}$. See text for the explanation of selected $z$-distances.
}
\label{fig:2}
\end{figure}

In Fig.~\ref{fig:2}(a) we show the numerically calculated intensity of the evolving soliton 
lattice. 
The two sublattices propagate in opposite directions and periodically collide, 
but they maintain their sublattice structures and propagation directions intact after every collision, 
which is ensured by the colliding properties of (Kerr-type) solitons~\cite{Chen2012}. 
The intercell and intracell distances are equal at $z=0$, because we have chosen $T=d/4$; 
$\kappa>0$ implies that the sublattices initially approach each other. Thus, 
in the $z$-interval from $z=0$ until the first collision, the soliton lattice has the structure 
of the trivial SSH lattice, see Fig.~\ref{fig:2}(a). 
After the first collision, the lattice retains its trivial topology until 
the intercell and intracell distances became equal again for the first time after $z=0$. 
This point is denoted with a vertical dashed line at $z=6.718$~mm in Fig.~\ref{fig:2}.
At that point, the lattice undergoes a topological phase transition from the trivial to the 
nontrivial SSH soliton lattice, illustrated in real space in Figs.~\ref{fig:2}(a) and \ref{fig:1}(c-left).

\begin{figure}
\includegraphics[width=0.45\textwidth]{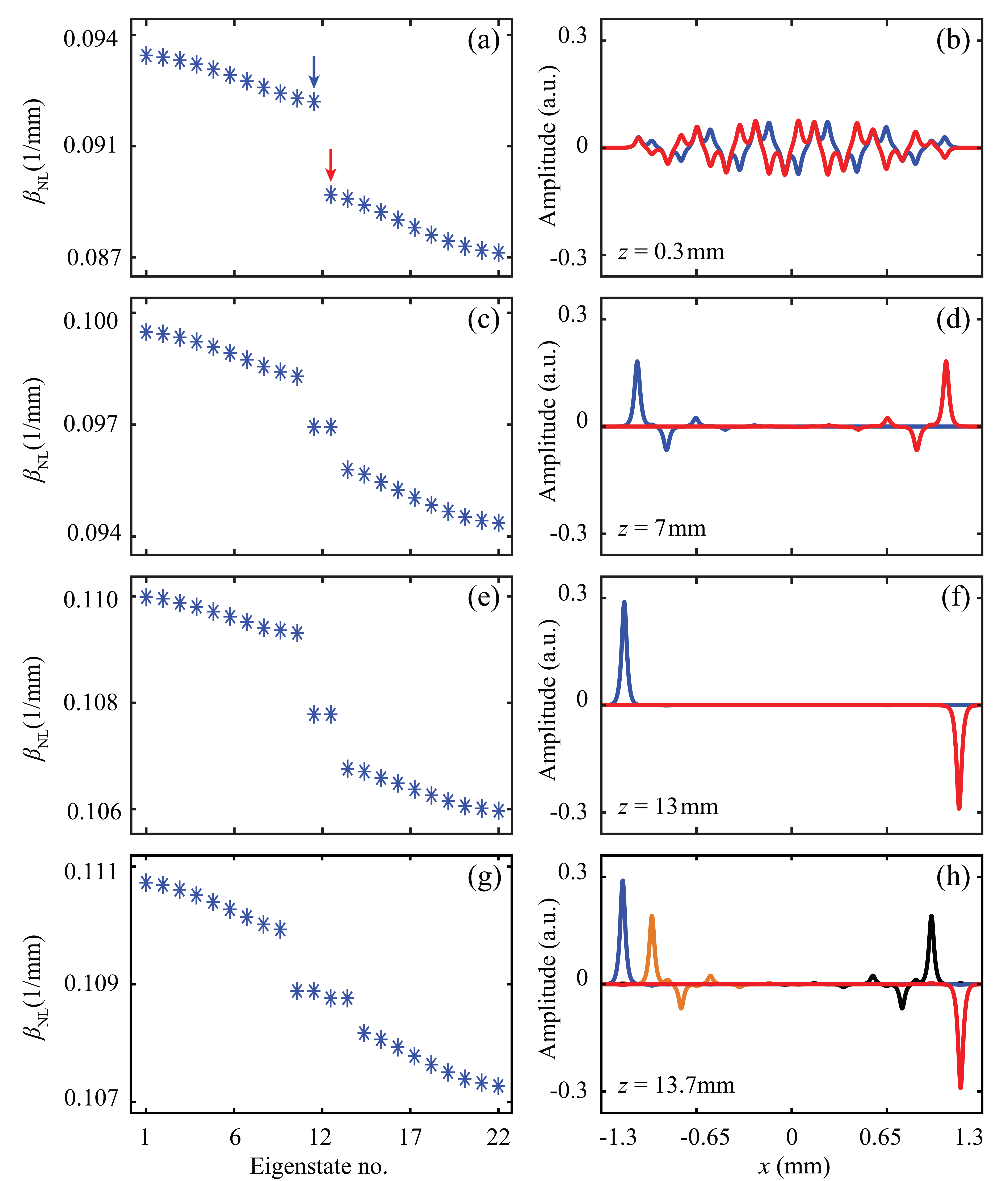}
\caption{Spectra of the evolving soliton lattice (left column) and selected eigenmodes $\phi_{NL,n}$
(right column) at propagation distances indicated by red arrows in Fig.~\ref{fig:2}.
(a) Spectrum and (b) two eigenmodes in the trivial phase at $z=0.3$~mm.
The two eigenmodes are closest to the gap as indicated with arrows in (a). 
(c) Spectrum and (d) topological localized states in the nontrivial phase at $z=7$~mm, 
just after the first topological phase transition.
(e) Spectrum and (f) localized states after the crossover from the 
nontrivial to the trivial phase at $z=13$~mm. The states are localized solely in the outermost solitons, 
their amplitude is zero in the bulk of the soliton lattice, which is in contrast 
from the amplitude-phase structure of topological edge states shown in (d). 
(g) Spectrum and (h) localized states at $z=13.7$~mm, 
after the 2nd phase transition. Two of the localized states are 
topological (black and orange lines), whereas the other two are outermost solitons (blue and red lines).}
\label{fig:3}
\end{figure}

An ultimate signature of the dynamical topological phase transition is illustrated in 
Fig.~\ref{fig:2}(b), which shows the band-gap structure of the evolving soliton lattice. 
We see that for $z$ values up to the first topological phase transition point at 
$z= 6.718$~mm, there are two bands without any states in the gap. 
At the transition point the gap closes and immediately re-opens, where two eigenvalues 
are pulled from the bands to stay within the gap.
These isolated eigenvalues correspond to the topologically nontrivial edge states of the 
SSH soliton lattice, with characteristic phase and amplitude structure, illustrated in 
Fig. \ref{fig:3}(d)~\cite{Xia2020, SSH, MalkovaOL2009}. 
They dynamically emerged at the transition point.
Gap closing is an inevitable and necessary ingredient of a topological phase transition that is clearly illustrated 
in Fig.~\ref{fig:2}(b). 

In order to fully unveil the behavior of our system, we explore the band gap 
structure and the modes of the SSH soliton lattice before the transition  
in Fig.~\ref{fig:3}(a,b) (for concreteness we consider $z=0.3$~mm), 
and just after the transition in Fig.~\ref{fig:3}(c,d) (at $z=7$~mm). 
At $z=0.3$~mm there are two bands separated by the gap, see Fig.~\ref{fig:3}(a).
All eigenmodes of the lattice are extended. In Fig.~\ref{fig:3}(b) we plot the two extended modes 
with eigenvalues closest to the gap. 
At the phase transition, these two extended modes are pulled from the band into 
the gap, see Fig.~\ref{fig:3}(c); at this point they became localized topological 
edge modes of the SSH soliton lattice, illustrated in Fig.~\ref{fig:3}(d). 
We see that both of them have the characteristic features of the topological edge modes: 
their amplitude is nonzero only in odd lattice sites (counting from the edge inwards), 
and the neighboring peaks in the mode amplitude are out of phase 
(see e.g. \cite{Xia2020, SSH, MalkovaOL2009}).

A glance at Fig.~\ref{fig:2}(b) shows an interesting feature of the evolving 
spectrum at $z=13.438$~mm: another gap closing and re-opening occurs, 
where two eigenstates bifurcate from the bands to become localized in the gap; 
see the inset in Fig.~\ref{fig:2}(b) and Fig.~\ref{fig:3}(g) and (h). 
This appears to be another topological phase transition from the trivial to the nontrivial 
SSH lattice. However, if this interpretation is correct (as we show in what follows), 
it means that the system is converted from the nontrivial to the trivial regime in between 
the two gap closing points shown in Fig.~\ref{fig:2}(b). 
This conversion is not a topological phase transition because 
the gap remains open at all propagation distances between $6.718$~mm and $13.438$~mm.

To explain this intriguing phenomenon, we need to resort to the real space dynamics in Fig.~\ref{fig:2}(a), 
and explore the region shaded in yellow where the soliton collisions take place. 
In this region two outermost solitons become separated from the lattice, because 
the distance to their nearest neighbors becomes $d$, which is the 
NNN distance in the SSH lattice, and thus the 
tunneling probability from these outermost solitons to the bulk of the SSH 
lattice is practically zero.
The eigenvalues corresponding to the outermost solitons are in the gap [see Fig.~\ref{fig:3}(e)], so 
the eigenmodes are obviously localized [see Fig.~\ref{fig:3}(f)], but their amplitude-phase 
structure does not possess the feature of topological edge states illustrated in Fig.~\ref{fig:3}(d). 
Thus, in the yellow region, two outermost solitons are actually decoupled from the SSH lattice, 
which leads to the crossover from the topologically nontrivial to the trivial phase. 
This crossover is fully equivalent to a gradual process of pulling two outermost lattice sites 
of the nontrivial SSH lattice into infinity, as illustrated in Fig.~\ref{fig:1}(d).

The existence of the crossover is in full agreement with the observation and interpretation 
of the gap closing point at $z=13.438$~mm in Fig.~\ref{fig:2}(b) described above. 
This pattern of alternating sequence of events - dynamical topological phase transitions 
(trivial-to-nontrivial phase) $\rightarrow$ crossover via decoupling of the outermost 
solitons (nontrivial-to-trivial phase), repeats itself during propagation until the two sublattices 
become separated and evolve without further collisions. 
The duration of this sequence depends on the number of solitons in each sublattice. 
The sublattice constant $d$ is chosen sufficiently large so that the NNN tunneling probability 
is negligible; therefore, when sublattices become separated, 
we can regard this system as a set of independent solitons.

In conclusion, we have found dynamical topological phase transitions in evolving nonlinear SSH soliton lattices, 
which are classified as emergent nonlinear topological phenomena, because they cease to exist if 
nonlinearity is turned off. 
These phase transitions convert the SSH soliton lattices from the topologically trivial-to-nontrivial phase, 
and are evinced by the gap closing and re-opening accompanied by emergence of two localized topological edge states. 
The eigenvalues of these edge states are pulled from the bands into the gap at the phase transition point. 
In addition, we have found a crossover from the topologically nontrivial-to-trivial regimes, which occurs via 
decoupling of the nontrivial edge states from the bulk of the lattice. 
These two opposing events occur one after the other in succession during the nonlinear dynamical evolution 
of the system. We have used the widely present Kerr-type nonlinearity to demonstrate our findings; 
it should be mentioned that in nonlinear saturable media such as photorefractive crystals, soliton collisions are 
typically not elastic (e.g., there can be fission and fusion of solitons), which means that direct observation 
of topological phase transitions proposed here should be more challenging (albeit not impossible) in saturable media. 
Nevertheless, we envisage this work will lead to exciting directions of fundamental research in nonlinear topological photonics.

This research is supported in part by the QuantiXLie Center of Excellence, a project co-financed by the Croatian Government and European Union through the European Regional Development Fund - the Competitiveness and Cohesion Operational Programme (Grant KK.01.1.1.01.0004), the National Key R\&D Program of China under Grant No. 2017YFA0303800, the National Natural Science Foundation (11922408, 91750204, 11674180), PCSIRT, and the 111 Project (No. B07013) in China. 
R.M. acknowledges support from the NSERC Discovery and the Canada Research Chair Programs. R.M. is affiliated to 5 as an adjoint faculty. 
D.B. acknowledges support from the 66 Postdoctoral Science Grant of China.
We acknowledge support from the Croatian-Chinese Scientific \& Technological Cooperation.


\end{document}